\documentclass[
aps,
prb,
twocolumn,
superscriptaddress,
longbibliography
]{revtex4-2}

\usepackage{amsfonts}
\usepackage{textcomp}
\usepackage{times}
\usepackage{graphicx}
\usepackage{float}
\usepackage{latexsym,amsmath,amssymb,bm,euscript}
\usepackage{color}
\usepackage{subfigure}
\usepackage{epstopdf}
\usepackage[colorlinks=true,linkcolor=blue,citecolor=blue]{hyperref}
\usepackage{soul}
\usepackage[normalem]{ulem}
\usepackage{mathrsfs}
\usepackage{amsmath}
\usepackage{xspace}
\usepackage{natbib}
\usepackage{ulem}
\usepackage{etoolbox}
\usepackage{tikz}
\usepackage{physics}
\usepackage{chemformula}
\usepackage{natbib}
\usepackage{tikz}
\usepackage{verbatim}

\newcommand{\hc}{\text{H.c.}}
\newcommand{\eff}{\text{eff}}
\newcommand{\gs}{\text{GS}}
\newcommand{\heff}{H_{\eff}}
\newcommand{\tpa}{t_{\parallel}}
\newcommand{\tpe}{t_{\perp}}
\newcommand{\zz}{\mathbb{Z}}
\newcommand{\oo}{\mathcal{O}}

\begin{document}

\title{Quantum Simulation with Gauge Fixing: from Ising Lattice Gauge Theory to Dynamical Flux Model}
\author{Junsen Wang}
\email{jsw@ucas.ac.cn}
\affiliation{Center of Materials Science and Optoelectronics Engineering, College of Materials Science and Opto-electronic Technology, University 
of Chinese Academy of Sciences, Beijing 100049, China}
\affiliation{CAS Key Laboratory of Theoretical Physics, Institute of Theoretical 
Physics, Chinese Academy of Sciences, Beijing 100190, China}

\author{Xiangxiang Sun}
\affiliation{Hefei National Research Center for Physical Sciences at the Microscale and School of Physical Sciences, University of Science and Technology of China, Hefei 230026, China}
\affiliation{CAS Center for Excellence in Quantum Information and Quantum Physics, University of Science and Technology of China, Hefei 230026, China}

\author{Wei Zheng}
\email{zw8796@ustc.edu.cn}
\affiliation{Hefei National Research Center for Physical Sciences at the Microscale and School of Physical Sciences, University of Science and Technology of China, Hefei 230026, China}
\affiliation{CAS Center for Excellence in Quantum Information and Quantum Physics, University of Science and Technology of China, Hefei 230026, China}
\affiliation{Hefei National Laboratory, University of Science and Technology of China, Hefei 230088, China}

\date{\today}

\begin{abstract}
Quantum simulation of synthetic dynamic gauge field has attracted much attentions in recent years. There are two traditional ways to simulate gauge theories. One is to directly simulate the full Hamiltonian of gauge theories with local gauge symmetries. And the other is to engineer the projected Hamiltonian in one gauge subsector. In this work, we provide the third way towards the simulation of gauge theories based on \emph{gauge fixing}. To demonstrate this concept, we fix the gauge of an Ising lattice gauge field coupled with spinless fermions on a ladder geometry. After the gauge fixing, this gauge theory is reduced to a simpler model, in which fermions hop on a ladder with a fluctuating dynamical $\zz_{2}$ flux. 
Then we show that this model can be realized via Floquet engineering in ultracold atomic gases. By analytical and numerical studies of this dynamical flux model, we deduce that there is a confinement-to-deconfinement phase transition in the original unfixed gauge theory. This work paves the way to quantum simulate lattice gauge theory using the concept of gauge fixing, relevant both for condensed matter and high energy physics.
\end{abstract}

\maketitle

\section{Introduction}

Lattice gauge theories (LGTs) have both fundamental and practical importance in modern physics~\cite{creutz1983,montvay1997,rothe2012}.
They are originally proposed by Wilson as a non-perturbative framework to deal with quantum chromodynamics (QCD) in the strong coupling region~\cite{Wilson1974}. Soon after, in combination with Monte-Carlo
methods, it becomes a standard numerical approach to QCD. 
LGT can also emerge from strongly correlated quantum materials, such as quantum
spin liquids~\cite{Broholm2020} and high $T_c$ superconductors~\cite{Lee2006}. More recently, the concept of LGT has extended to the territory of quantum computation and information. For example, the celebrated toric code model is essentially a $\mathbb{Z}_2$ LGT. However due to the massive local constraints imposed by the gauge symmetry and the limitation of classical computers, it is challenging to study the real-time dynamics of LGTs. 

In the last decade, quantum simulation based on artificial quantum systems, such as ultracold atoms~\cite{wiese2013,zhou2022,mildenberger2022,wang2023,zhang2023}, trapped ions~\cite{martinez2016,davoudi2020,bazavan2023}, Rydberg atoms in optical tweezers~\cite{surace2020,domanti2023,cheng2024}, and superconducting circuits~\cite{byrnes2006,muschik2017,zohar2017,bender2018,klco2018,klco2020,mathis2020}, gradually evolves as a refreshing tool to attack
this hard problem~\cite{tagliacozzo2013,zohar2015,bermudez2015,assaad2016,gonzalezCuadra2017,smith2017,brenes2018,smith2018,zache2018,yao2020,zheng2020,halimeh2020,vanDamme2020,halimeh2020fate,halimeh2021,zohar2021,cheng2022,gao2022,halimeh2022,halimeh2022enhancing,gao2023,osborne2023,homeier2023,halimeh2023,halimeh2023robust,van2023,kebri2024,su2024,qi2024,tang2024}. Up to now there are two main routes to quantum simulate LGT. The first is to directly simulate the full Hamiltonian of LGT with massive local gauge symmetries. For example, in 2019, a $\mathbb{Z}_{2}$ LGT has been realized via Floquet engineering in a double well~\cite{schweizer2019}. In 2020, a quantum link model, one particular U(1) LGT, has been simulated up to 71 sites in an optical lattice~\cite{yang2020}. In the same year, U(1) LGT has also been realized in atomic mixture~\cite{mil2020}. The second route is to simulate the Hamiltonian in one gauge sector. In this situation, the local gauge symmetry can emerge from the local constraints. For instance, the Rydberg blockade effect has been used to simulate the quantum link model in the gauge invariant subsector~\cite{surace2020,cheng2024}. These progresses have motivated extensive studies in the two directions.

In this work, we propose \emph{gauge fixing} as the third route towards quantum simulation of LGTs. In fact, gauge fixing has been widely used to deal with gauge theories defined in continuous space-time. For example in electromagnetism, one can fix the gauge by choosing the Coulomb, Lorenz or Landau gauge in practical calculations. Though gauge fixing is not obligatory in LGTs, it has been implemented in the context of Monte-Carlo computation of LGTs~\cite{Creutz1983a}. However, gauge fixing is much less explored in the modern era of quantum simulation on the Hamiltonian level. Compare to the usual Faddeev-Popov-De Witt gauge fixing procedure based on path integral~\cite{faddeev1967,dewitt1967}, we clarify the concept of gauge fixing on the Hamiltonian level. The proper gauge fixing procedure require that the matrix elements of both the Hamiltonian and gauge invariant observables to be unchanged. After gauge fixing, the rigorous requirement of local gauge invariance is bypassed, and the Hilbert space is significantly reduced. The fixed Hamiltonian becomes much simpler and relatively easy to implement in experiment. 

To illustrate the gauge fixing procedure without any further complications, we fix the gauge of an Ising LGT coupled with fermions on a ladder geometry, which is the \emph{minimal and simplest} model for lattice gauge field coupled with matter field featuring plaquette interactions. After gauge fixing, the Hamiltonian is largely simplified, and describes fermions hopping on a ladder subject to a fluctuating dynamical flux. We then propose a Floquet engineering scheme to simulate this gauge-fixed Hamiltonian.  
We used two species ultracold fermions on a ladder optical lattice. One is to simulate the gauge field, and the other to simulate the matter field. The zero temperature phase diagram of this fixed model is determined via analysis in the limiting cases and the numerical density-matrix renormalization group (DMRG) calculation. We note that this model basically exhibit two phases. One is the antiferromagnetic N\'eel order phase, and the other is the paramagnetic phase, which are reminiscent of the deconfinement and confinement phases in the original unfixed model.

This paper is organized as follows. In the next section, we first clarify the concept of gauge fixing on the Hamiltonian level. Then in Sec.~\ref{sec:fixing}, we fix the gauge of an Ising LGT coupled with fermions on a ladder geometry. We then propose a Floquet engineering scheme to simulate this gauge-fixed Hamiltonian in Sec.~\ref{sec:floquet}. Next the ground state of this model is determined via analysis in the limiting cases and the numerical DMRG calculations, in Sec.~\ref{sec:limiting-cases} and \ref{sec:dmrg}, respectively. Lastly, we give a discussion and outlook in Sec.~\ref{sec:discussion-outlook}.

\begin{figure*}[t!]
    \centering
    \includegraphics[width=\textwidth]{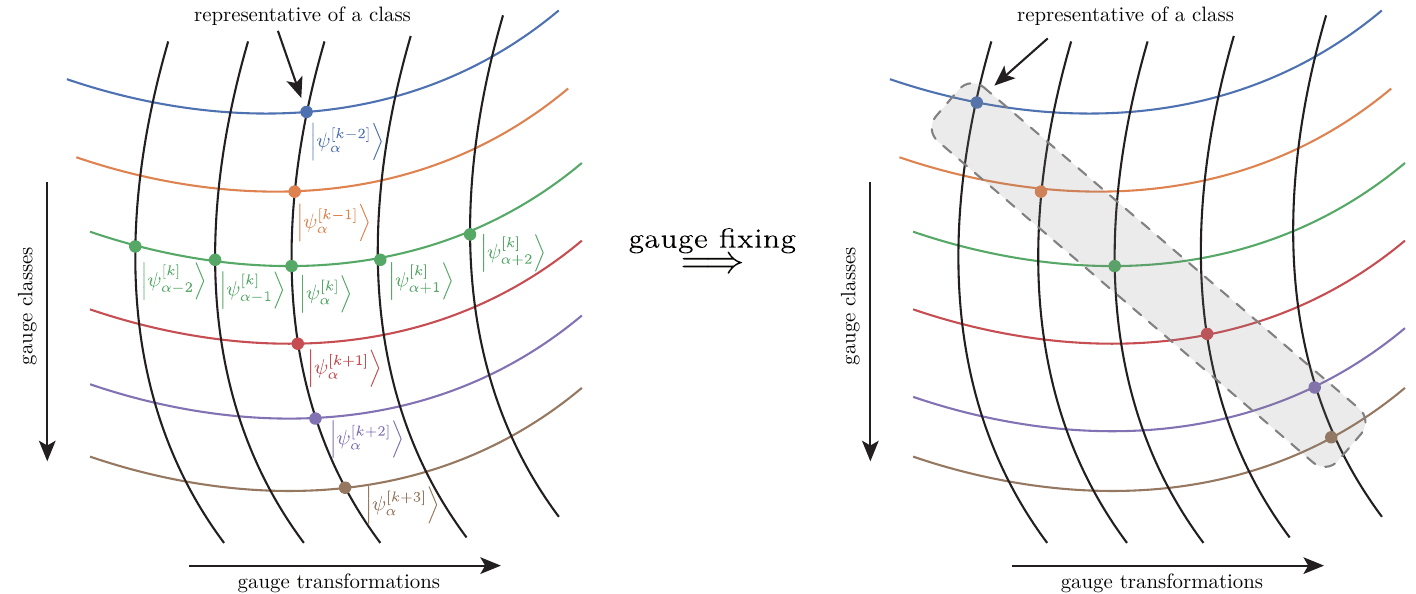}
    \caption{(a) States are classified into different gauge classes (represented by different horizontal lines with different colors). In each class, different states are related by a gauge transformations (represented by different vertical lines). (b) Gauge fixing amounts to picking up one particular state in each class, as shown by the gray band enclosing one state in each gauge class.}
    \label{fig:gauge_fixing}
\end{figure*}

\section{Gauge fixing on the Hamiltonian level}

Gauge fixing is crucial for gauge theories in continuous space-time. The Feynman integral of the historical paths connected by gauge transformation will give a divergent propagator. Thus in continuous space-time, one has to choose a particular gauge to eliminate the redundancy of gauge transformation. Say, the original path integral for a gauge theory is given by 
\begin{equation}
    Z = \int DA\, e^{iS(A)}.\label{eq:I}
\end{equation}
Now we introduce the resolution of identity, $1 = \Delta(A) \int DG \, \delta[f(A_g)]$, where $f$ is some function of our choice, $A_G$ is the gauge field under gauge transformation $G$, and $\Delta(A)$ is the so-called Faddeev-Popov determinant \cite{faddeev1967}, and $\delta [...]$ is a Dirac delta functional. Note that the integral is over the group of gauge transformation $G$.
Eq.~\eqref{eq:I} then can be rewritten as
\begin{align}
    Z &= \int DA\, e^{iS(A)} \Delta(A) \int DG\, \delta[f(A_G)]\\
    &= \left(\int DG \right) \int DA\, e^{iS(A)} \Delta(A) \delta[f(A)],
\end{align}
where in the second line we interchange the order of integration, then followed by changing $A$ to $A_{G^{-1}}$ and using the fact that $DA$, $S(A)$ and $\Delta(A)$ are all invariant under gauge transformation, $A\rightarrow A_{G^{-1}}$. In this way the divergent group integration $(\int DG)$ has been factored out, and as an overall coefficient, it is simply thrown away.

In LGTs, gauge fixing is not essential. However, just as discussed in introduction, gauge fixing can largely reduce the Hilbert space, thus can effectively simplify the experiments for simulations. In the following we will first introduce the concept of gauge fixing on the Hamiltonian level.

Given a Hamiltonian of a LGT, ${H}$. It possesses local gauge symmetry as $[{G}_{i},{H}]=0$, where ${G}_{i}$ is the local gauge transformation operator defined on each site. As a result, ${G}_{i}$ and ${H}$ share the eigen states,
\begin{eqnarray}
H\left| \psi  \right\rangle  = E\left| \psi  \right\rangle , \\ 
G_i \left| \psi  \right\rangle  = g_i \left| \psi  \right\rangle .
\end{eqnarray}
According to $g_{i}$, the eigen values of ${G}_{i}$, the Hamiltonian can be block diagonalized into disconnected gauge sectors, 
\begin{eqnarray}
H = \left[ {\begin{array}{*{20}c}
    \ddots  & {} & {} & {}  \\
   {} & { H_{\left\{ {g'_i } \right\}} } & {} & {}  \\
   {} & {} & { H_{\left\{ {g_i } \right\}} } & {}  \\
   {} & {} & {} &  \ddots   \\
 \end{array} } \right].
\end{eqnarray}
Here ${H}_{\left\{g_{i}\right\}}$ is the Hamiltonian inside each gauge sector.

Usually people are interested in the so-called \emph{physical sector} or the \emph{gauge invariant sector}, in which $g_{i}=1$ for all site. Thus all states in this sector are gauge invariant, $ G_i \left| \psi_{\mathrm{GI}}  \right\rangle  =  \left| \psi_{\mathrm{GI}}  \right\rangle$.

All orthogonal wave functions can be classified into different gauge classes (see Fig.~\ref{fig:gauge_fixing}),  
\begin{eqnarray}
C^{\left[ 1 \right]}  = \left( {\left| {\psi _1^{\left[ 1 \right]} } \right\rangle ,\left| {\psi _2^{\left[ 1 \right]} } \right\rangle , \cdots ,\left| {\psi _V^{\left[ 1 \right]} } \right\rangle } \right), \\ 
  C^{\left[ 2 \right]}  = \left( {\left| {\psi _1^{\left[ 2 \right]} } \right\rangle ,\left| {\psi _2^{\left[ 2 \right]} } \right\rangle , \cdots ,\left| {\psi _V^{\left[ 2 \right]} } \right\rangle } \right), \\ 
   \vdots  \\ 
  C^{\left[ D \right]}  = \left( {\left| {\psi _1^{\left[ D \right]} } \right\rangle ,\left| {\psi _2^{\left[ D \right]} } \right\rangle , \cdots ,\left| {\psi _V^{\left[ D \right]} } \right\rangle } \right),
\end{eqnarray}
where $V$ is the number of total independent gauge transformations, and $D$ is the number of gauge classes. Wave functions belong to the same gauge class are related by a gauge transformation, $\left| {\psi _\alpha ^{\left[ k \right]} } \right\rangle  = G_{\alpha\beta}^{[k]} \left| {\psi _\beta ^{\left[ k \right]} } \right\rangle$. The wave functions belong to different gauge classes can not be transformed into each other by any gauge transformations. 

The basis of the gauge invariant sector can be constructed by these gauge classes via equal-weight superposition of all the wave functions inside one gauge class, 
\begin{eqnarray}
\left| {\psi _{{\text{GI}}}^{\left[ k \right]} } \right\rangle  = \frac{1}
{{\sqrt V }}\sum\limits_{\alpha  = 1}^V {\left| {\psi _\alpha ^{\left[ k \right]} } \right\rangle }.
\end{eqnarray}
Such superposition is gauge invariant, since any gauge transformation is just a rearrangement of the summation. The gauge fixing on the Hamiltonian level is freezing some degree of freedoms of the gauge field. The goal of the freezing is to pick up one particular wave function in each gauge class, 
\begin{eqnarray}
C^{\left[ 1 \right]} \xrightarrow{{{\text{gauge fixing}}}}\left| {\psi _\alpha ^{\left[ 1 \right]} } \right\rangle , \\ 
  C^{\left[ 2 \right]} \xrightarrow{{{\text{gauge fixing}}}}\left| {\psi _\beta ^{\left[ 2 \right]} } \right\rangle , \\ 
   \vdots  \\ 
  C^{\left[ D \right]} \xrightarrow{{{\text{gauge fixing}}}}\left| {\psi _\gamma ^{\left[ D \right]} } \right\rangle ,
\end{eqnarray}
After this gauge fixing the dimension of the Hilbert space is suppressed to $D$, and the gauge transformations can not be applied. At the same time, the gauge fixing rule is also applied on the Hamiltonian and gauge invariant observables,
\begin{eqnarray}
  H\xrightarrow{{{\text{gauge fixing}}}} H_{{\text{fixed}}} , \\ 
  O\xrightarrow{{{\text{gauge fixing}}}} O_{{\text{fixed}}},
\end{eqnarray}
We require the gauge fixing rule to ensure that 
\begin{eqnarray}
     \left\langle {\psi _{{\text{GI}}}^{\left[ q \right]} } \right|H\left| {\psi _{{\text{GI}}}^{\left[ p \right]} } \right\rangle {\text{ = }}\left\langle {\psi _\alpha ^{\left[ q \right]} } \right|H_{{\text{fixed}}} \left| {\psi _\beta ^{\left[ p \right]} } \right\rangle , \\ 
  \left\langle {\psi _{{\text{GI}}}^{\left[ q \right]} } \right|O\left| {\psi _{{\text{GI}}}^{\left[ p \right]} } \right\rangle {\text{ = }}\left\langle {\psi _\alpha ^{\left[ q \right]} } \right|O_{{\text{fixed}}} \left| {\psi _\beta ^{\left[ p \right]} } \right\rangle. 
\end{eqnarray}
Note that after the gauge fixing, one faithfully reconstruct the matrix elements of Hamiltonian in the gauge invariant sector.
Moreover, the gauge-fixing approach is equally able to study real-time dynamics of LGTs.
Note, of course, the quantities accessible must be gauge-invariant ones.
For example, the gauge fixing rules ensure that the real-time two-point gauge-invariant correlator becomes
\begin{eqnarray}
      && \langle O_1(t) O_2(0) \rangle \\
      &=& \left\langle {\psi _{{\text{GI}}}^{\left[ q \right]} } \right| e^{i H t} O_1 e^{-i H t} O_2 \left| {\psi _{{\text{GI}}}^{\left[ p \right]} } \right\rangle\\
      &=& \left\langle {\psi _\alpha ^{\left[ q \right]} } \right| e^{i H_{\text{fixed}} t} O_{{1,\text{fixed}}} e^{-i H_{\text{fixed}} t} O_{2,\text{fixed}} \left| {\psi _\beta ^{\left[ p \right]} } \right\rangle,
\end{eqnarray}
and all real-time higher-point gauge-invariant correlators can be similarly obtained.

If one wants to calculate the eigen states in the gauge invariant sector, we can first compute the eigen states of $H_{{\text{fixed}}} $ as 
\begin{equation}
\left| \psi  \right\rangle  = \sum\limits_{q = 1}^D {a_q \left| {\psi _\alpha ^{\left[ q \right]} } \right\rangle },
\end{equation}
Then the corresponding gauge invariant eigen state can be expressed as
\begin{eqnarray}
  \left| {\psi _{{\text{GI}}} } \right\rangle  &=& \frac{1}
{{\sqrt V }}\sum\limits_{\beta  = 1}^V { G_\beta  } \left| \psi  \right\rangle  \\ 
   &=& \sum\limits_{q = 1}^D {a_q \frac{1}
{{\sqrt V }}\sum\limits_{\beta  = 1}^V { G_\beta  } \left| {\psi _\alpha ^{\left[ q \right]} } \right\rangle }  \\ 
   &=& \sum\limits_{q = 1}^D {a_q \left| {\psi _{{\text{GI}}}^{\left[ q \right]} } \right\rangle }, 
\end{eqnarray}
where ${G_\beta}$ are all possible independent gauge transformations. In the gauge fixing process, some degree of freedoms are frozen. We have to unfreeze them when reconstructing the gauge invariant wave functions, such that the gauge transformation can be applied. 

\section{\label{sec:fixing}Gauge fixing for an Ising lattice gauge field coupled with fermions on a ladder}

\begin{figure*}[t!]
    \centering
    \includegraphics[width=.6\textwidth]{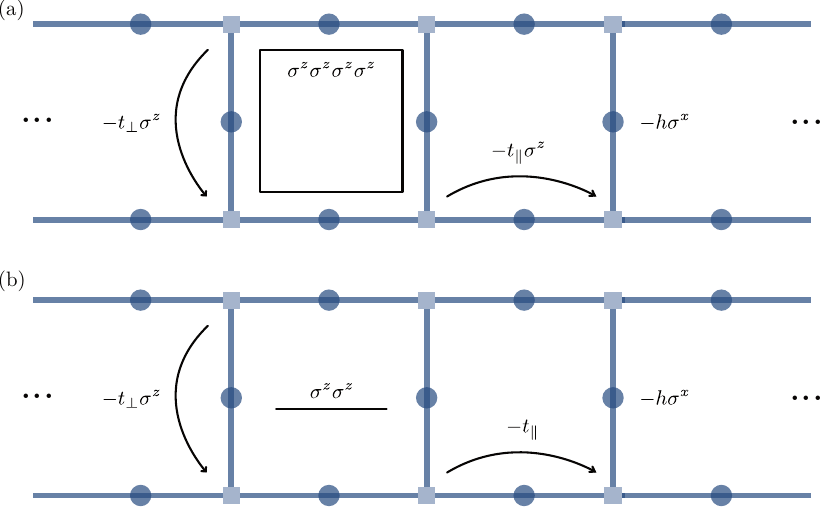}
    \caption{(a) Schematic of the gauge invariant Hamiltonian, Eq.~\eqref{eq:hgi}. Spinless fermions live on the vertices and Ising spins live on the legs and rungs. (b) Schematic of the gauge invariant Hamiltonian \emph{after gauge fixing}, Eq.~\eqref{eq:hgf}. Note that the Ising spins now only live on the rungs.}
    \label{fig:latt}
\end{figure*}

As a minimal and simplest example, we consider the gauge invariant Hamiltonian of an Ising lattice gauge field coupled with spinless fermions, defined on a ladder geometry [see Fig.~\ref{fig:latt}(a)], 
\begin{align}
  H &=  - t_ \bot  \sum\limits_j {\left( {c_{jL}^\dag  \sigma _j^z c_{jR}  + {\text{H}}{\text{.c}}{\text{.}}} \right)} \nonumber \\ 
   &\quad -t_\parallel  \sum\limits_{j\alpha} {\left( {c_{j\alpha}^\dag  \sigma _{ \left\langle {j,j + 1} \right\rangle \alpha }^z c_{j+1, \alpha}  + {\text{H}}{\text{.c}}{\text{.}}} \right)} \nonumber \\ 
   &\quad + U_1\sum\limits_j {\sigma _j^z \sigma _{\left\langle {j,j + 1} \right\rangle, R}^z \sigma _{j + 1}^z \sigma _{\left\langle {j,j + 1} \right\rangle,L}^z }  - h\sum\limits_j {\sigma _j^x },\label{eq:hgi}
\end{align}
Here, the first two lines describe fermions hopping along the longitudinal (leg) and transverse (rung) directions interacting with the Ising gauge fields according to the minimal coupling procedure. And  
${\sigma _{\alpha \left\langle {j,j + 1} \right\rangle }^z }$ denotes the Ising gauge field sitting on the $\alpha=L,R$ leg between site $j$ and site $j+1$. The last line is the $\mathbb Z_2$ analog of magnetic field and electric field energy, respectively. 
Comparing to the standard Ising LGT, here the electric field term only exists on the rungs. 

The Hamiltonian Eq.~\eqref{eq:hgi} has local Ising gauge symmetries. The corresponding gauge transformation operator is 
\begin{equation}
G_{j\alpha}  = e^{i\pi n_{j\alpha} } \sigma _{ \left\langle {j - 1,j} \right\rangle,\alpha }^x \sigma _j^x \sigma _{ \left\langle {j,j + 1} \right\rangle,\alpha }^x,
\end{equation}
where $n_{j\alpha }  = c_{j\alpha }^\dag  c_{j\alpha} $. It commutes with the Hamiltonian,
\begin{equation}
  \label{eq:4}
[{G}_{j\alpha}, {H}]=0,
\end{equation}
As we discussed in the previous section, in gauge invariant sector, ${G}_{j\alpha}=1$ on all site. As a result, in this sector we have 
\begin{equation}
e^{i\pi n_{j\alpha} }  = \sigma _{ \left\langle {j - 1,j} \right\rangle,\alpha }^x \sigma _j^x \sigma _{ \left\langle {j,j + 1} \right\rangle,\alpha }^x.
\end{equation}
This is nothing but the $\mathbb{Z}_{2}$ version of Gauss law, which imposes extensive local constrains on the dynamics of the system. 

In a gauge theory, only the gauge invariant observable has non-vanishing expectation values. In our model there are two common gauge invariant observables. One is the Wilson loop, 
\begin{equation}
W  \left( {i,j} \right) = \expval{\sigma _i^z \left( {\prod\limits_{k = i}^{j - 1} {\sigma _{\left\langle {k,k + 1} \right\rangle,L }^z \sigma _{\left\langle {k,k + 1} \right\rangle,R }^z } } \right)\sigma _j^z} ,\label{eq:wgij1}
\end{equation}
which can be used to diagnose confinement or deconfinement phases. The other is the gauge invariant correlations of fermions,
\begin{equation}
C_{\alpha \alpha } \left( {i,j} \right) = \expval{c_{i\alpha }  \left( {\prod\limits_{{\text{string}}} {\sigma ^z } } \right)c^\dag_{j\alpha}},\label{eq:cabij1}
\end{equation}
where the string connects the two fermion sites

\begin{figure*}[!t]
    \includegraphics[width=\textwidth]{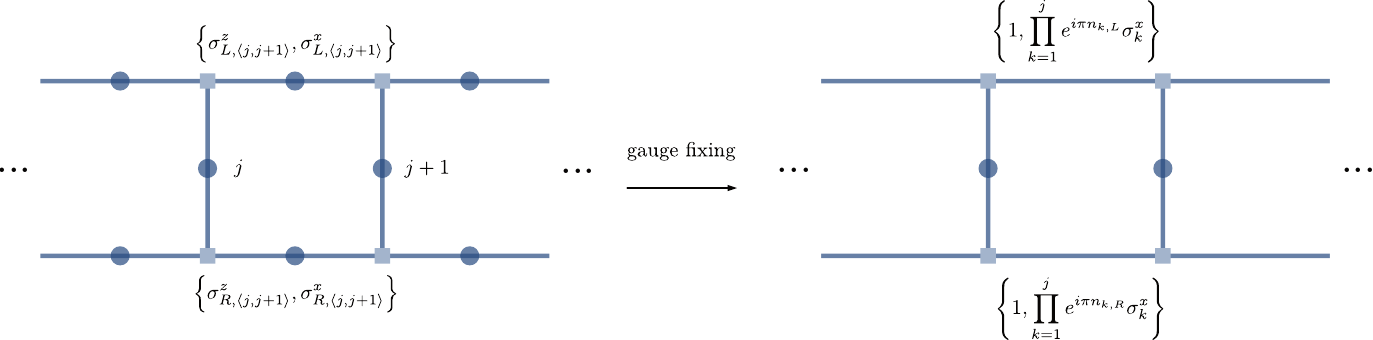}
    \caption{Gauge fixing for an Ising lattice gauge field coupled with fermions on a ladder. Before the gauge fixing, shown in the left, there is a spin-1/2 (fermionic) degree of freedom on each link (site), denoted by a blue dot (square). After gauge fixing, shown in the right, local Hilbert spaces on the links of two legs are eliminated, indicated by the absence of blue dots and the replacement of the corresponding operators there. One can consecutively fix the gauge for two legs starting from the leftmost side.}
    \label{fig:gf}
\end{figure*}

Now we are going to fix the gauge of this model. A typical basis of this Ising Lattice gauge model is 
\begin{eqnarray}
  &&\left| \psi  \right\rangle  = \left| \varphi  \right\rangle _{{\text{fermi}}}  \hfill \nonumber\\
  &&\quad  \otimes \left| {\xi _1 ,\xi _2 , \cdots ;\xi _{\left\langle {1,2} \right\rangle,L } ,\xi _{\left\langle {2,3} \right\rangle,L } , \cdots ;\xi _{\left\langle {1,2} \right\rangle,R } ,\xi _{\left\langle {2,3} \right\rangle,R } , \cdots } \right\rangle , 
  \hfill  \nonumber\\
\end{eqnarray}
where $\xi  =  \uparrow , \downarrow $ is the states of the gauge field on the link. Then one can perform the gauge fixing process by freezing the gauge field degree of freedom on the legs as 
\begin{equation}
\left| \psi  \right\rangle \xrightarrow{{{\text{gauge fixing}}}}\left| \varphi  \right\rangle _{{\text{fermi}}}  \otimes \left| {\xi _1 ,\xi _2 , \cdots ; \uparrow , \uparrow , \cdots ; \uparrow , \uparrow , \cdots } \right\rangle ,
\end{equation}
Then the gauge fixing of the spin operator on the legs corresponds to 
\begin{equation}
\sigma _{ \left\langle {i,j + 1} \right\rangle ,\alpha}^z \xrightarrow{{{\text{gauge fixing}}}}1,
\end{equation}
Since $\sigma _{\left\langle {j,j + 1} \right\rangle ,\alpha }^x $ on the legs can flip the flux of the corresponding plaquette, it can not be simply replaced by a c-number. Instead, one can use the Gauss law to replace it by other spin operators as 
\begin{equation}
\sigma _{ \left\langle {j,j + 1} \right\rangle,\alpha }^x  = e^{i\pi \hat n_{j\alpha } } \sigma _{ \left\langle {j - 1,j} \right\rangle,\alpha }^x \sigma _j^x ,
\end{equation}
Moreover, $\sigma _{\left\langle {j - 1,j} \right\rangle,\alpha  }^x $ should also be fixed. Thus we can obtain 
\begin{equation}
\sigma _{\left\langle {j,j + 1} \right\rangle,\alpha  }^x  = \prod\limits_{k = 1}^j {e^{i\pi \hat n_{k\alpha } } \sigma _k^x } ,
\end{equation}
Note that it becomes a highly non-local operators. 
So our gauge fixing rules for the spin operators on legs are summarized as 
\begin{eqnarray}
  \sigma _{ \left\langle {j,j + 1} \right\rangle,\alpha }^z &\xrightarrow{{{\text{gauge fixing}}}}& 1, \\ 
  \sigma _{ \left\langle {j,j + 1} \right\rangle,\alpha }^x &\xrightarrow{{{\text{gauge fixing}}}}&\prod\limits_{k = 1}^j {e^{i\pi \hat n_{ k\alpha} } \sigma _k^x } \label{sigmaxonleg}
\end{eqnarray}
After this gauge fixing process, the Hamiltonian becomes~[see Fig.~\ref{fig:latt}(b)]
\begin{eqnarray}
  H_{{\text{fixed}}}  &=&  - t_ \bot  \sum\limits_j {\left( {c_{jL}^\dag  \sigma _j^z c_{jR}  + {\text{H}}{\text{.c}}{\text{.}}} \right)} \nonumber \\ 
   &&- t_\parallel  \sum\limits_{ j\alpha} {\left( { c_{j\alpha }^\dag  c_{j + 1,\alpha}  + {\text{H}}{\text{.c}}{\text{.}}} \right)}  \nonumber \\ 
   &&+ U_1\sum\limits_j {\sigma _j^z \sigma _{j + 1}^z }  - h\sum\limits_j {\sigma _j^x },
   \label{eq:hgf}
\end{eqnarray}
Note that in our simple model, there is no electric field term on the legs. Therefore after gauge fixing the Hamiltonian is still local. The Wilson loop and fermion correlation becomes
\begin{eqnarray}
   W_\Gamma  \left( {i,j} \right) &\xrightarrow{{{\text{gauge fixing}}}} & \expval{\sigma _i^z \sigma _j^z} ,\label{eq:wgij} \\ 
  {C}_{\alpha \alpha } \left( {i,j} \right) &\xrightarrow{{{\text{gauge fixing}}}} & \expval{c_{i\alpha} c^\dag_{j\alpha} },\label{eq:cabij}
\end{eqnarray}
Note that the Wilson loop becomes a two-point correlator. After gauge fixing the model loses the local gauge invariance, and the dimension of the Hilbert space is largely reduced. This gauge fixing process is illustrated in Fig.~\ref{fig:gf}.

\begin{figure*}[!t]
  \centering
  \includegraphics[width=0.65\textwidth]{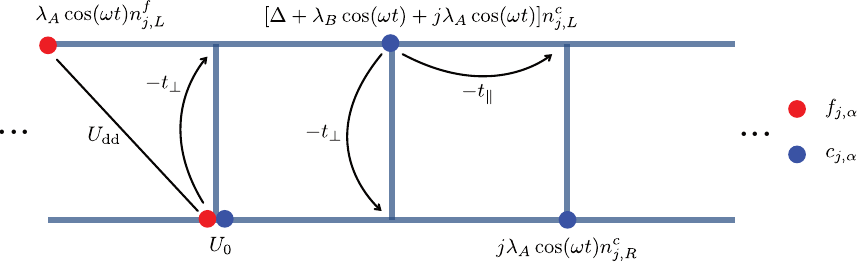}
  \caption{Floquet engineering scheme. Red and blue points represents $f$ and $c$ fermions, respectively. The former can only hop along the rung direction (and each rung contains exactly one $f$ femrion), while the latter can hop along both directions. There is an on-site interaction between them, and a dipole-dipole interaction between $f$ fermions. Three time-dependent on-site potential $V_i(t)$, $i=1,2,3$, are added. The first two are only applied to the left leg, while the last one is applied to both legs for the $c$ fermion, with the amplitude increasing linearly from left to right.}
  \label{fig:fe}
\end{figure*}

\section{\label{sec:floquet}Floquet engineering of the Hamiltonian with fixed gauge}

We now discuss how to use a simple Floquet engineering scheme to realize the gauge-fixed Hamiltonian, Eq.~\eqref{eq:hgf}. Consider an optical lattice forming a ladder geometry, loaded with two types of fermions, $c$ alkali atoms and $f$ alkaline earth atoms with large magnetic dipole moment or polar molecules with large electric dipole moment, as shown in Fig.~\ref{fig:fe}. The optical lattice is species dependent such that the $c$ fermion can hop along both the leg and the rung directions, while the $f$ fermions can only hop along the rung direction. Thus the hopping term of the Hamiltonian reads
\begin{align}
    H_0 &=-t_{\perp}\sum_j(f_{jL}^{\dagger}f_{jR}+c_{jL}^{\dagger}c_{jR})\nonumber\\
        &\quad -t_{\parallel}\sum_{j\alpha}c_{j+1,\alpha}^{\dagger}c_{j\alpha} + \hc, \label{eq:h0}
\end{align}
where $L$ ($R$) denotes the left (right) leg of the ladder. We further assume that the system is carefully prepared, and \emph{only one $f$ fermion is loaded in each rung}, i.e.,
\begin{equation}
    n_{j L}^f + n_{j R}^f = 1, \quad \forall j. \label{eq:nf}
\end{equation}
There is an on-site interaction between two kinds of fermions, with strength $U_0$,
\begin{align}
  H_I &= U_{0}\sum_{j\alpha}n_{j\alpha}^fn_{j\alpha}^c.\label{eq:hi}
\end{align}
A strong uniform magnetic field or an electric field is applied perpendicular to the plane of the ladder to polarize the $f$ fermions. Thus the dipole moments of all $f$ fermions are frozen to the perpendicular direction. As a result, the long range dipole-dipole interaction between the $f$ fermions is
\begin{equation}
    H_{{\text{dd}}}  = \sum\limits_{\alpha \beta ,ij} {U_{\mathrm{dd}}\left( {{\mathbf{r}}_{i \alpha}  - {\mathbf{r}}_{j\beta} } \right)n_{i\alpha}^f n_{j\beta}^f },
    \label{eq:hdd}
\end{equation}
where $U_{{\text{dd}}} \left( {\mathbf{r}} \right) = C_{{\text{dd}}} /4\pi \left| {\mathbf{r}} \right|^3 $. By tuning the lattice spacing, one can ignore the dipole-dipole interaction beyond nearest neighboring rungs. Since there is only one $f$ fermion on each rung, the spacing between $f$ fermions on the nearest neighboring rungs has twofold values, the lattice constant in the leg direction $a$ and diagonal of the plaquette $\sqrt{a^2+b^2}$, where $b$ is the lattice constant in the rung direction. Then Eq.~\eqref{eq:hdd} can be simplified into
\begin{equation}
H_{{\text{dd}}}  = U_{\text{1}} \sum\limits_{\alpha j} {\left( {n_{j\alpha}^f n_{j+1,\alpha}^f  - n_{j\alpha}^f n_{j+1,\bar \alpha}^f } \right)}
\end{equation}
where $U_{\text{1}}  = \left[ {U_{{\text{dd}}} \left( a \right) - U_{{\text{dd}}} \left( {\sqrt {a^2  + b^2 } } \right)} \right]/2$, and a constant is ignored. 

We then modulate the on-site potential periodically, which consists of three parts, $V(t)=V_1(t)+V_2(t)+V_3(t)$, where
\begin{subequations}
  \label{eq:9}
  \begin{align}
      V_1(t) &= \sum_j \lambda_A\cos(\omega t)n_{jL}^f, \label{eq:v1}\\
  V_2(t) &= \sum_j[\Delta+\lambda_B\cos(\omega t)]n_{jL}^c, \label{eq:v2}\\
  V_3(t) &= \sum_jj\lambda_A\cos(\omega t)(n_{jR}^c+n_{jL}^c).\label{eq:v3}
  \end{align}
\end{subequations}
Namely, on the left leg only there are periodically driven on-site potential $V_{1,2}(t)$ for both types of fermions, with different amplitudes, $\lambda_{A,B}$, and a relative energy offset $\Delta$. For the $c$ fermion, there is an additional periodically driven gradient potential $V_{3}(t)$ along both legs. The full Hamiltonian is then given by $H=H_0+H_I+H_{\text{dd}}+V(t)$. The energy offset and the driving frequency are turned such that
\begin{equation}
   \omega = \Delta = U_0 / 2 \gg U_1, t_{\perp}, t_{\parallel},
\end{equation}

\begin{figure}[!t]
  \centering
  \includegraphics[width=0.46\textwidth]{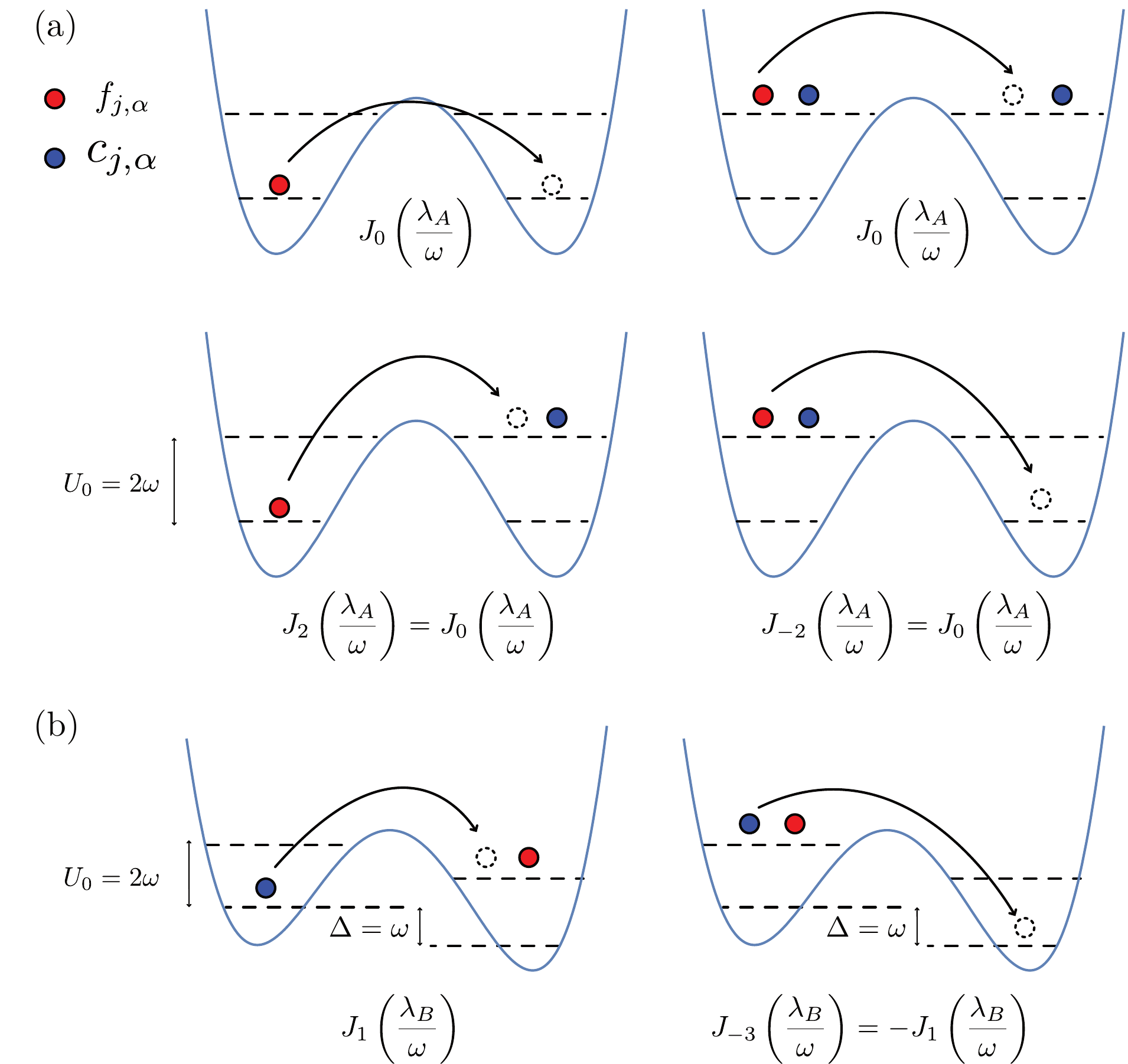}
  \caption{The hopping process of $f$ fermions in the double well can be understood in (a). If both sides have no $c$ fermion or have one, the hopping amplitude is the same, as shown by the first row. If $f$ fermion hops from a side without (with) $c$ fermion to the other side with (without) $c$ fermion, it will absorb (emit) two photons. By tuning the potential amplitude $\lambda_A$, this hopping amplitude can be made the same as the previous case. The hopping of $c$ fermions in the transverse double well can be understood in (b). In this case, due to the the relative energy offset between two sides, if $c$ fermion hops from a side without (with) $f$ fermion to the other side with (without) $f$ fermion, it will obsorb (emit) one (three) photons. By tuning the potential amplitude $\lambda_B$, these two hopping amplitudes can be made with same magnitude but with \emph{opposite sign}. The hopping process of $c$ fermions in the longitudinal direction has a similar mechanism as (a).}
  \label{fig:fei}
\end{figure}

We now derive the time-independent, effective Hamiltonian of the system. Via a unitary transformation,
\begin{equation}
  R (t) = e^{i \int_0^t d \tau [H_I + V (\tau)]}, \label{ut}
\end{equation}
the fermion operator $b_{j \alpha}$, for $b = f$ or $c$, generally transforms in the following way,
\begin{equation}
R(t)b_{j\alpha}R^{\dagger}(t)=[(1-n_{j\alpha}^{\bar{b}})+n_{j\alpha}^{\bar{b}}e^{-i2\omega t}]e^{-i\theta_{j\alpha}^a(t)}b_{j\alpha},
\end{equation}
where $\bar{b} = f$ ($c$) if $b = c$ ($f$) and
\begin{align}
  \theta_{jL}^f(t) &= \frac{\lambda_A}{\omega}\sin(\omega t),\\
  \theta_{jR}^f(t) &= 0,\\
  \theta_{jL}^c(t) &= \omega t+\frac{j\lambda_A+\lambda_B}{\omega}\sin(\omega t),\\
  \theta_{jR}^c(t) &= \frac{j\lambda_A}{\omega}\sin(\omega t).
\end{align}
I.e., the fermion operator acquires both a time-dependent operator factor and a numerical
phase factor. The former, resulting from the interaction in Eq.~\eqref{ut}, depends
on the population of the other kind of fermion, while the latter can be viewed as a gauge transformation. Indeed, one trades off the time-dependent scalar potential (electric field) in Eq.~\eqref{eq:9} by a time-dependent vector
potential, appearing as a phase due to the Peierls substitution. The full
Hamiltonian in the rotating frame then becomes
\begin{align}
    H_R(t) &\equiv R(t)H(t)R^{\dagger}(t)-R(t)i\partial_{t}R^{\dagger}(t) \\
           &= \{-t_{\perp}\sum_j[\hat{A}_j^f(t)e^{i\delta\theta_j^f(t)}f_{jL}^{\dagger}f_{jR}\nonumber\\
  &\quad +\hat{A}_j^c(t)e^{i\delta\theta_j^c(t)}c_{jL}^{\dagger}c_{jR}] \nonumber\\
  &\quad -t_{\parallel}\sum_{j\alpha}[\hat{A}_j^{\alpha}(t)e^{i\delta\theta_j^{\alpha}}c_{j+1,\alpha}^{\dagger}c_{j\alpha}]+\hc \}+H_{\mathrm{dd}},\label{hu}
\end{align}
where each hopping term contains an additional time-dependent operator, given respectively by
\begin{align}
  \hat{A}_j^f(t) &= (1-n_{jL}^c)(1-n_{jR}^c)+n_{jL}^cn_{jR}^c\nonumber\\
  &\quad +(1-n_{jL}^c)n_{jR}^ce^{-i2\omega t}+n_{jL}^c(1-n_{jR}^c)e^{i2\omega t},\\
  \hat{A}_j^c(t) &= (1-n_{jL}^f)n_{jR}^fe^{- i 2 \omega t} +(1-n_{fR}^f)n_{jL}^fe^{i2\omega t},\label{eq:13}\\
  \hat{A}_j^{\alpha}(t) &= \left({1-n_{j+1,\alpha}^f}\right)(1-n_{j\alpha}^f)\nonumber\\
                 &\quad +n_{j+1,\alpha}^fn_{j\alpha}^f+(1-n_{j+1,\alpha}^f)n_{j\alpha}^fe^{-i2\omega t}\nonumber\\
  &\quad +n_{j+1,\alpha}^f(1-n_{j\alpha}^f)e^{i2\omega t},
\end{align}
where we have used Eq.~\eqref{eq:nf} to simplify Eq.~\eqref{eq:13}. The phase differences
shown in Eq.~\eqref{hu} are
\begin{align}
    \delta\theta_j^f(t) &= \theta_{jL}^f(t)-\theta_{jR}^f(t)=\frac{\lambda_A}{\omega}\sin(\omega t),\\
  \delta\theta_j^c(t) &= \theta_{jL}^c(t)-\theta_{jR}^c(t)=\omega t+\frac{\lambda_B}{\omega}\sin(\omega t),\\
  \delta\theta_j^{\alpha}(t) &= \theta_{j+1,\alpha}^c-\theta_{j\alpha}^c=\frac{\lambda_A}{\omega}\sin(\omega t),\quad \alpha=L,R.
\end{align}
Up to this point, everything is exact. Now consider the case with $\omega \gg t_{\parallel}, t_{\perp}$, which enables us to perform a high-frequency expansion on the rotated Hamiltonian Eq.~\eqref{hu}. Using the Jacobi-Anger identity, $e^{iz\sin\phi}=\sum_{n=-\infty}^{\infty}J_n(z)e^{in\phi}$, the effective Hamiltonian at the leading order becomes
\begin{align}
  H_{\eff} &\approx [-t_{\perp}\sum_j(\hat{B}_j^ff_{jL}^{\dagger}f_{jR}+\hat{B}_j^cc_{jL}^{\dagger}c_{jR})\nonumber\\
  &\quad -t_{\parallel}\sum_{j\alpha}\hat{B}_j^{\alpha}c_{j+1,\alpha}^{\dagger}c_{j\alpha}+\hc] \nonumber \\ & \quad+
  U_{\text{1}} \sum\limits_{j} (n_{jL}^f-n_{jR}^f) (n_{j+1,L}^f-n_{j+1,R}^f),
\end{align}
where the operator associated with each hopping term now respectively becomes
\begin{subequations}\label{eq:15}
  \begin{align}
  \hat{B}_j^f &= J_0\left(\frac{\lambda_A}{\omega}\right)[(1-n_{jL}^c)(1-n_{jR}^c)+n_{jL}^cn_{jR}^c]\nonumber\\
  &\quad +J_2\left(\frac{\lambda_A}{\omega}\right)[(1-n_{jL}^c)n_{jR}^c+n_{jL}^c(1-n_{jR}^c)],\\
  \hat{B}_j^c &= J_1\left(\frac{\lambda_B}{\omega}\right)(1-n_{jL}^f)n_{jR}^f-J_3\left(\frac{\lambda_B}{\omega}\right)(1-n_{fR}^f)n_{jL}^f,\\
  \hat{B}_j^{\alpha} &= J_0\left(\frac{\lambda_A}{\omega} \right) \left[\left({1-n_{j+1,\alpha}^f}\right)(1-n_{j\alpha}^f)+n_{j+1,\alpha}^fn_{j \alpha}^f \right]\nonumber\\
  &\quad +J_2\left(\frac{\lambda_A}{\omega}\right)[(1-n_{j+1,\alpha}^f)n_{j \alpha}^f+n_{j+1,\alpha}^f(1-n_{j\alpha}^f)],
\end{align}
\end{subequations}
where we have used the property of the Bessel function of the first kind that
$J_{-m}(z)=(-1)^mJ_m(z)$. Finally, setting
\begin{equation}
  J_2\left(\frac{\lambda_A}{\omega}\right)=J_0\left(\frac{\lambda_A}{\omega}\right), \quad J_3\left(\frac{\lambda_B}{\omega}\right)=J_1\left(\frac{\lambda_B}{\omega}\right),
  \label{eq:lalb}
\end{equation}
Eq.~\eqref{eq:15} is then simplified dramatically,
\begin{align}
    \hat{B}_j^f &= \hat{B}_j^{\alpha}=J_0\left(\frac{\lambda_A}{\omega}\right),\\
  \hat{B}_j^c &= J_1\left(\frac{\lambda_B}{\omega}\right)(n_{jR}^f-n_{fL}^f).
\end{align}
Note the minimum $\lambda_A/\omega$ and $\lambda_B/\omega$ that satisfy Eq.~\eqref{eq:lalb} are approximately $1.84$ and $3.05$.
Physically speaking, by tuning the lattice modulation amplitude $\lambda_A$,
the hopping process of one type of fermions can become no longer sensitive to the
population of the other kind, as schematically illustrated in Fig.~\ref{fig:fei}(a).
This mechanism works for both the $f$ fermion, and the $c$ fermions hopping along
the legs. We further utilize the property that the Bessel function of the first
kind satisfies $J_m (z) = - J_{- m} (z)$, for $m$ odd, to effectively flips
the sign of hopping strength for $c$ fermions when hop along the rung,
\emph{depending on the population of the $f$ fermion}, as illustrated in Fig.~\ref{fig:fei}(b).
Finally, we define the local spin operator [note that the $f$ fermion number at each rung is fixed to one, cf. Eq.~\eqref{eq:nf}]
\begin{subequations}
    \begin{align}
        \sigma_j^z &= n_{jL}^f-n_{jR}^f,\\
        \sigma_j^x &= f_{jL}^{\dagger}f_{jR}+f_{jR}^{\dagger}f_{jL},\\
        \sigma_j^y &= -i(f_{jL}^{\dagger}f_{jR}-f_{jR}^{\dagger}f_{jL}),
    \end{align}
\end{subequations}
which amounts to defining the two-dimensional local Hilbert space at each rung, with spin-up states corresponding to the fermion on the left side of the rung, and spin-down state corresponding to the fermion on the right side.
It then follows that the effective Hamiltonian takes the following compact form:
\begin{align}
  \heff &= -\tilde{t}_{\parallel}\sum_{j\alpha}\left(c_{j\alpha}^{\dagger}c_{j+1,\alpha}+\hc\right)\nonumber\\
  &\quad -\tilde{t}_{\perp}\sum_j\left(c_{jL}^{\dag}\sigma_j^z c_{jR}+\hc\right)\nonumber\\
  &\quad -h\sum_j\sigma_j^x + U_{1}\sum\limits_j {\sigma _j^z \sigma _{j + 1}^z } ,
  \label{eq:h}
\end{align}
where
\begin{align}
    \tilde{t}_{\parallel} &= t_{\parallel}J_0\left(\frac{\lambda_A}{\omega}\right),\\
  \tilde{t}_{\perp} &= t_{\perp}J_1\left(\frac{\lambda_B}{\omega} \right),\\
  h &= t_{\perp}J_0\left(\frac{\lambda_A}{\omega}\right).
\end{align}
Note that the long range dipole-dipole interaction maps to the Ising interaction between the spins on the rungs. This Hamiltonian is precisely the gauge-fixed Hamiltonian we discussed in the last section, Eq.~\eqref{eq:hgf}. Hereafter, we will drop the tilde symbol for parameters used in Eq.~\eqref{eq:h} for simplicity.

This effective Hamiltonian has fermion charge $\text{U}(1)$ symmetry, generated by $U_c(\phi) = e^{i\phi n}$, where $n=\sum_{j\alpha} n_{j\alpha}$ and $[n,\heff] = 0$.
There is another global $\text{U}(1)$ symmetry generated by $R_x (\phi) = e^{i \phi S_x}$, where $S^x = \frac{1}{2}\sum_j \left( c_{j L}^{\dag} c_{j R} + \hc \right)$ and $[S^x,\heff]=0$.
Moreover, Eq.~\eqref{eq:h} also has a discrete global $\mathbb{Z}_2$ symmetry generated by one of the following operators,
\begin{equation}
	Q_{\alpha} = \prod_j e^{i \pi n_{j\alpha}} \sigma_j^x,\qfor{\alpha = L,R}.
	\label{eq:gz2}
\end{equation}
These two operators are related by a global $\text{U}(1)$ rotation, $Q_L Q_R = U_c (\pi)$.
Also note that $Q_{\alpha}^2 = 1$, hence their eigenvalues are $\pm 1$.
Under this $\mathbb{Z}_2$ symmetry transformation, we have
\begin{equation}
	\sigma_j^z \rightarrow Q_{\alpha}\sigma_j^z Q_{\alpha}^{- 1} = - \sigma_j^z, \quad \forall j.
	\label{eq:sz}
\end{equation}
Hence, a nonzero ground state expectation value of $\sigma_j^z$ corresponds to
spontaneously breaking this global $\mathbb{Z}_2$ symmetry.
The system also has a particle-hole (PH) symmetry at half filling, namely, under the unitary
PH transformation, $c_j \rightarrow (-1)^j c_j^{\dagger}$ and $c_j^{\dagger}
\rightarrow (-1)^j c_j$, the Hamiltonian is invariant.
Note this PH symmetry not only swaps particles and holes, i.e., $N_c \rightarrow 2 N - N_c$, but also flips
the sign of the global $\mathbb{Z}_2$ charge $Q_{\alpha} \rightarrow (-1)^N Q_{\alpha}$ if the total number of rungs $N$ is odd.

\section{\label{sec:limiting-cases}Analytical results in the limiting cases}

In this section, we study two limiting cases of our effective Hamiltonian Eq.~\eqref{eq:h}. It is found that, for the electric field strength $h$ sufficiently small, the system is in the gapped phase with spontaneously broken global $\mathbb Z_2$ symmetry. In the large $h$ limit, the system also remains charge gapped at half-filling.

\subsection{\label{subsec:smallh}The small $h$ limit: Peierls instability}

We start with the $h=U_1=0$ case, where quantum fluctuations of the Ising fields are turned off, hence all $\sigma_i^z$ are conserved quantities and the full Hilbert space becomes a direct sum of different Ising spin configurations.
In each sector, the model describes free fermions on a ladder, with the hopping strength along two legs being ${t}_{\parallel}$, and along rungs being $\pm{t}_{i,\perp}$, where the $\pm$ sign corresponds to the eigenvalue of $\sigma_i^z$.
This Hamiltonian can be solved easily via exact diagonalization.
And we find numerically that, at half filling, the lowest energy among all sectors corresponds to a N\'eel-type antiferromagnetic ordering for these Ising spins.
This ground state spontaneously breaks the translational symmetry, and each plaquette contains a $\pi$ flux, which resembles the dimerized lattice distortion that underlies the fermionic Peierls instability at half filling~\cite{Peierls2001}. Adding a small positive $U_1$ does not modify this picture, since the spin-spin Ising interaction is also anti-ferromagnetic.

\begin{figure}[!t]
  \centering
  \includegraphics[width=0.46\textwidth]{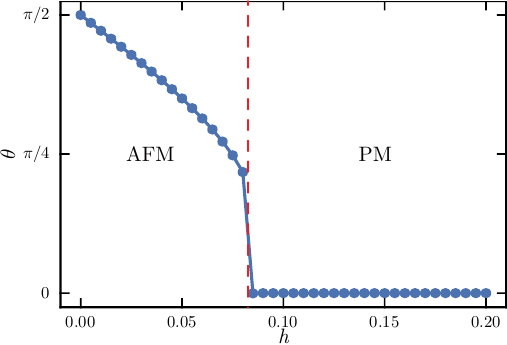}
  \caption{Rotation angle $\theta$ as a function of transverse field $h$ obtained by minimizing the ground state energy $\epsilon_{\gs}$, Eq.~\eqref{eq:gse}, for fixed $\tpe = 0.1$ and $U_1 = 0.2$. The red dashed line is a guide to the eye to indicate the approximate location of the phase transition point.}
  \label{fig:thetah}
\end{figure}

To elaborate on this analogy, and also to go to a finite but small $\zz_2$ electric field strength ${h}\ll {t}_{\parallel,\perp}$, we consider a Born-Oppenheimer-type approximation~\cite{GonzalezCuadra2019}.
Namely, we assume that the fermions adapt instantaneously to the Ising spin background.
Thus the latter provides an effective potential energy for the former, which in turn determines the ground-state spin configuration.
This amounts to using the following variational wavefunction ansatz
\begin{equation}
  \ket{\Psi_{\gs}} = \ket{\psi^{f}_\gs}\otimes e^{-i\sum_i \frac{\theta_{i}}{2}\sigma^y_{i}}\ket{\uparrow\uparrow\cdots\uparrow}_{x},
  \label{eq:vwf}
\end{equation}
where $\ket{\psi^{f}_{\gs}}$ is the fermion wavefunction, say, in the local site Fock basis, and $\ket{\uparrow\uparrow\cdots\uparrow}_x$ means all spins pointing to the $+x$ direction.
For fermions at half filling, the numerical result has showed the doubling of the unit cell for the ground state, we thus use a single variational angle $(\theta_{2i-1},\theta_{2i})= (\theta,-\theta)$ for the Ising spins due to reflection symmetry.
Then the ground state energy can be written as an analytical function of $\theta$, $\tpe$, $h$ and $U_1$ (we set $\tpa$ as the energy unit):
\begin{align}
 \epsilon_{\gs}(\theta) &= 8\sqrt{4+\tpe^2\sin^2\theta} \mathrm E\left (\frac{4}{4+\tpe^2\sin^2\theta}\right )\nonumber\\
 &\quad - h \cos\theta - U_1 \sin^2\theta
 \label{eq:gse}
\end{align}
where $\mathrm E(x):= \int_{0}^{\pi/2}\sqrt{1-x \sin^2\theta}d\theta$ is the complete elliptic integral of the second kind. One can then straightforwardly minimize this ground state energy for a fixed $\tpe$, $h$ and $U_1$ to find the optimal $\theta$. In Fig.~\ref{fig:thetah}, we plot the numerical results for $\theta$ as a function of $h$ obtained by minimizing the ground state energy $\epsilon_{\gs}$ with fixed $\tpe=0.1$ and $U_1=0.2$. It shows that under this Born-Oppenheimer-type approximation, there is a first-order quantum phase transition between the symmetry-breaking antiferromagentic phase to the disordered paramagnetic phase.

Recall the Peierls instability of the SSH model~\cite{Su1979}, where the energy reduction of the fermions due to a gap opening is compensated by the increase of elastic energy of the lattice distortion~\cite{Peierls2001}.
Here, the $\zz_2$ electric field term, $-h\cos\theta$, plays the role of elastic energy.
The central difference in this analogy is that while the one-dimensional fermion chain is always unstable towards a dimerization due to Peierls instability; here, for large $h$, the mechanism does not work. This is due to the fact that our assumption, the Born-Oppenheimer-type argument, fails, i.e., Eq.~\eqref{eq:vwf} becomes inappropriate.

\subsection{The large $h$ limit: repulsive Hubbard model}

In the infinite $h$ limit, the system becomes two decoupled fermion chains, which is gapless.
A large but finite $h$ introduces coupling between these two chains.
To examine the fate of this infinite-$h$ gapless phase under such perturbations, in the following, we will derive the effective Hamiltonian for Eq.~\eqref{eq:h} in the large $h$ limit, then phases of the resulting Hamiltonian is discussed. Since $U_1$ is small in our experimental scheme, in the following discussion we will focus on the case with $U_1=0$, and it is expect that the conclusion holds for a small nonzero $U_1$.

We start by writing the original Hamiltonian Eq.~\eqref{eq:h} with $U_1=0$ as $H = h H_0 + H_1 + V$, where
\begin{align}
  H_0 &= - \sum_i \sigma_i^x,\\
  H_1 &= - t_{\parallel} \sum_{\langle i j \rangle, \tau} \left( c_{i
  \tau}^{\dag} c_{j \tau} + \text{H.c.} \right),\\
  V &= - t_{\perp} \sum_i \left( c_{i, L}^{\dag} \sigma_i^z c_{i, R} +\hc \right).
\end{align}
Upon a canonical and unitary transformation, $\ket{\psi'} = e^{i S} \ket{\psi}$, where $S$ is Hermitian and time-independent, it becomes (using the Baker-Campbell-Hausdorff formula),
\begin{equation}
  \tilde{H} = e^{i S} He^{- iS} = H + i [S, H] + \frac{i^2}{2!} [S, [S, H]] + \cdots.
  \label{eq:hp}
\end{equation}
We write the operator $S$ as a series expansion in $h^{- 1}$,
\begin{equation}
  S = S^{(1)} h^{- 1} + S^{(2)} h^{- 2} + \cdots = \sum_{l = 1}^{\infty}
   S^{(l)} h^{- l} .
\end{equation}
Up to the zeroth order in $h^{- 1}$, Eq.~\eqref{eq:hp} reads
\begin{equation}
  H' = h H_0 + H_1 + V + i [S^{(1)}, H_0] +\mathcal{O} (h^{- 1}) .
  \label{eq:hp0}
\end{equation}
By choosing $S^{(1)} = \frac{t_{\perp}}{2} \sum_i \left( \sigma^y_i c_{i L}^{\dag} c_{iR} +\hc \right)$, we have $V + i [S^{(1)}, H_0] = 0$.
And the effective Hamiltonian Eq.~\eqref{eq:hp0} after the projection, $PH' P$, where $P$ projects to the low-energy subspace with all spins pointing to the $+ \hat{x}$ direction, becomes (omitting the constant)
\begin{equation}
  \tilde{H}^{(1)} = - t_{\parallel} \sum_{\langle i j \rangle \tau} \left(
   c_{i \tau}^{\dag} c_{j \tau} +\hc \right) +\mathcal{O} (h^{- 1}),
\end{equation}
which is just two free fermion chain, as expected. Up to the first order in $h^{- 1}$, Eq.~\eqref{eq:hp} becomes
\begin{align}
  H' &= h H_0 + H_1\nonumber\\
     &\quad + (i [S^{(1)}, H_1] + i [S^{(1)}, V] + i [S^{(2)}, H_0])
       h^{- 1}\nonumber\\
  &\quad +\mathcal{O} (h^{- 2}).
  \label{eq:hp1}
\end{align}
where the first commutator reads
\begin{equation}
  i [S^{(1)}, H_1] = - \frac{it_{\perp} t_{\parallel}}{2} \sum_{\langle i j \rangle \tau}
  (\sigma_i^y - \sigma_j^y) (c_{i \tau}^{\dag} c_{j \bar{\tau}} - c_{j
  \tau}^{\dag} c_{i \bar{\tau}}),
\end{equation}
with $\bar{\tau} = L$ $(R)$ if $\tau = R$ $(L)$. The second one is found to be $i [S^{(1)}, V] = t_{\perp}^2 \sum_i \sigma_i^x (2 n_{i L} n_{i R} - n_i)$, with $n_i = n_{i L} + n_{i R}$.
By choosing $S^{(2)}$ to eliminate the off-diagonal terms in the $\sigma^x$ basis, namely, demanding that $i [S^{(1)}, H_1] + i [S^{(2)}, H_0] \stackrel{!}{=} 0$, we then have
\begin{equation}
  S^{(2)} = - \frac{it_{\perp} t_{\parallel}}{4} \sum_{\langle i j \rangle
   \tau} (\sigma_i^z - \sigma_j^z) (c_{i \tau}^{\dag} c_{j \bar{\tau}} - c_{j
   \tau}^{\dag} c_{i \bar{\tau}}).
\end{equation}
And Eq.~\eqref{eq:hp1} becomes (with constant terms omitted)
\begin{align}
  \tilde{H}^{(2)} &= - t_{\parallel} \sum_{\langle i j \rangle \tau} \left(
    c_{i \tau}^{\dag} c_{j \tau} +\hc \right)\nonumber\\
  &\quad + \frac{2
   t_{\perp}^2}{h} \sum_i \sigma_i^x n_{i L} n_{i R} +\mathcal{O} (h^{-
    2}).
\end{align}
In the large $h$ limit, spins are polarized and are hard to fluctuate. Thus, one can replace $\sigma^{x}_i$ by its expectation value, $\sigma^{x}_i\approx\expval{\sigma^{x}_i}\approx1$. This replacement is equivalent to projecting the Hamiltonian into the low-energy manifold. Then we obtain
\begin{align}
  \tilde{H}^{(2)} &\approx - t_{\parallel} \sum_{\langle i j \rangle \tau} \left(
    c_{i \tau}^{\dag} c_{j \tau} +\hc \right)\nonumber\\
  &\quad + \frac{2
   t_{\perp}^2}{h} \sum_i n_{i L} n_{i R} +\mathcal{O} (h^{-
    2}).
\end{align}
Note that this low energy effective Hamiltonian is equivalent to the one-dimensional Fermi Hubbard model with the repulsive interaction strength $U=2t_{\perp}^2/h$, and hopping strength $t=t_{\parallel}$. The leg index $\tau = L,R$, plays the role of spin index. 

\begin{figure}[!t]
\includegraphics{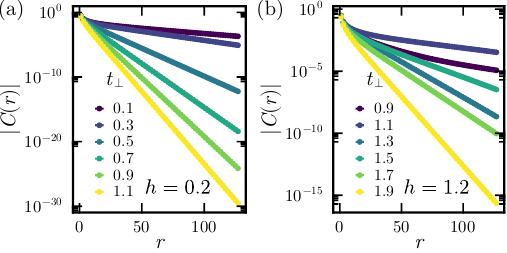}
\caption{\label{lsccdag} Behavior of the gauge-invariant single-particle correlator of fermion matter, defined in Eq.~\eqref{eq:ccdag}, for (a) $h = 0.2$ and (b) $h=1.2$. $t_\parallel = 1$ is set as the energy unit, and $U_1=0.1$.}
\end{figure}

The quantum phase diagram of the 1D repulsive Fermi Hubbard model as a function of $U/t$ is well-known. For a  given commensurate filling and increasing $U/t$, there is a quantum phase transition from the Luttinger liquid (metallic) phase to the Mott insulator phase of the Berezinskii–Kosterlitz–Thouless type.
For half-filling, the transition point is at $U/t = 0$~\cite{Giamarchi1997}. That means the system will stay in the Mott phase at any finite on-site interaction strength. 
Combining with the analysis obtained from the small $h$ limit, we conclude that the charge gap should always be open for the half-filling case at arbitrary finite $h$

\section{\label{sec:dmrg}Numerical results: A DMRG study}

We now present the numerical results. We use the density matrix renormalization group algorithm \cite{white1992} based on matrix product states \cite{schollwoeck2011} via the ITensor Library \cite{Fishman2022}, to study the full phase diagram of the gauge-fixed model Eq.~\eqref{eq:h} at half filling.
We emphasize that, behaviors of various gauge invariant quantities of the gauge-invariant Hamiltonian Eq.~\eqref{eq:hgi} can be examined from this gauge-fixed one.
In the following simulations, $t_\parallel$ is fixed to be the energy unit, and we fix $U_1=0.1$. We choose the open boundary conditions with total rungs of ladder to be $L=256$. The fermion number $\text{U}(1)$ symmetry is used, with maximal bond dimension $D = 256$ and truncation error $\lesssim 10^{-6}$.

\subsection{Fermion matter sector}

We start with the fermion matter sector. First of all, in the half-filling case, it is found that the (gauge-invariant) charge density distribution is always uniform, with the corresponding density-density correlator always decaying exponentially.
We then turn to the gauge-invariant single-particle correlator, defined in Eq.~\eqref{eq:cabij1}, which explicitly reads
\begin{eqnarray}
C_{LL}(r) &=& \expval{ c_{i,L} \pqty{\prod_{j=i}^{i+r-1} \sigma^z_{jL}} c^\dagger_{i+r,L} }\nonumber\\
&&\xrightarrow{{{\text{gauge fixing}}}} \expval{ c_{i,L} c^\dagger_{i+r,L} }.
\label{eq:ccdag}
\end{eqnarray}
The behavior of this quantity for different $\tpe$ is shown in Fig.~\ref{lsccdag}, one always finds an exponential decay, which is consistent with our theoretical analysis given in Sec.~\eqref{sec:limiting-cases}. Namely, the charge gap remains open for all $h$.

\begin{figure}[!t]
\includegraphics{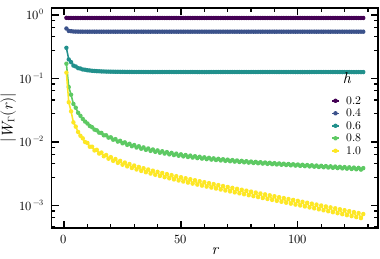}
\caption{\label{lsszsz} Behavior of the Wilson loop $\tilde{W}_{\Gamma}(r)$, Eq.~\eqref{eq:wgij}, for various $h$ with $\tpe = \tpa = 1$ and $U_1=0.1$.}
\end{figure}

\subsection{Ising gauge sector}

One of the most important gauge-invariant observables in Ising gauge sector is the Wilson loop~\cite{Kogut1979}.
In its original context, the confinement-deconfinement transition can be defined by the area-law or perimeter-law decay of this quantity for a sufficiently large loop~\cite{Shankar2017}. For a ladder geometry, it is defined in Eq.~\eqref{eq:wgij1}. After the gauge-fixing process, it reduces to a two-point correlator, Eq.~\eqref{eq:wgij}. In Fig~\ref{lsszsz}, for fixed $\tpa = \tpe = 1$ and $U_1=0.1$, we find that it quickly converges to a finite value for small $h$, which is reminiscent of a deconfined phase of the $\zz_2$ gauge field. While in the disordered phase for large $h$, the Wilson loop $|W_{\Gamma}(r)|$ decays exponentially to zero at large distances, which is reminiscent of a confined phase of the $\zz_2$ gauge field.

\begin{figure}[!t]
\includegraphics{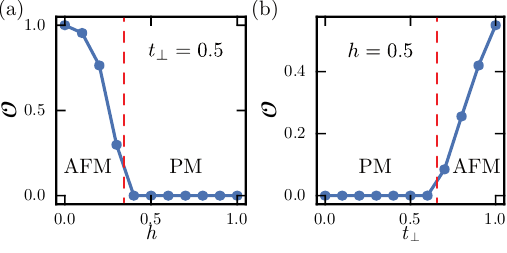}
\caption{\label{fig:lsz2}Behavior of the order parameter $\oo$, Eq.~\eqref{eq:o}, as a function of (a) $h$ with fixed $t_\perp = 0.5$ and (b) $\tpe$ with fixed $h = 0.5$. $t_\parallel = 1$ is set as the energy unit, and $U_1=0.1$. The red dashed lines are a guide to the eye to indicate the approximate location of the phase transition point.}
\end{figure}

\begin{figure}[t]
\includegraphics{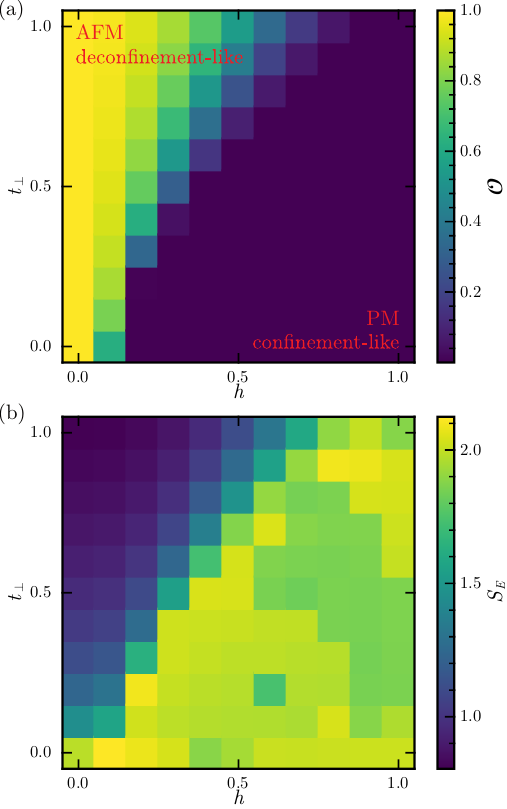}
\caption{\label{fig:matz2}(a) Contour plot of the order parameter $\mathcal{O}$. (b) Contour plot of the entanglement entropy $S_E$ measured at the central site. The other parameters: $\tpa=1$ and $U_1=0.1$.}
\end{figure}

\subsection{Global $\zz_2$ symmetry breaking and full phase diagram}

The effective Hamiltonian Eq.~\eqref{eq:h} has a global $\zz_2$ symmetry generated by Eq.~\eqref{eq:gz2}. As discussed in Sec.~\ref{subsec:smallh}, a Peierls instability mechanism will lead to a N\'eel-type ordering along spin-$z$ direction for small $h$. Thus this symmetry is spontaneously broken there. By increasing $h$, these Ising spins will tend to aligned along spin-$x$ direction. It is therefore expected that there is a order-disorder phase transition, in analog to the antiferromagnetic transverse field Ising chain.
To identify the phase transition point explicitly, we use the order parameter defined by
\begin{equation}
    \oo = \frac{1}{N}\sum_i (-1)^i \expval{\sigma_i^z}.
    \label{eq:o}
\end{equation}
In Fig.~\ref{fig:lsz2}, this order parameter is plotted as a function of $h$ with fixed $\tpe=0.5$, and as a function of $\tpe$ with fixed $h=0.5$. It is found that a sufficiently large $h$ will kill the symmetry-breaking phase, while a sufficiently large $\tpe$ will induce this symmetry-breaking phase. We then present the full phase diagram in Fig.~\ref{fig:matz2}(a). There are basically two phases,  for large $\tpe$ and small $h$, the system is in the antiferromagnetic (AFM) phase, while for $h$ large and $\tpe$ small, the system is in the paramagnetic phase (PM). In the AFM phase, the Wilson loop quickly converges to a finite value, which is a deconfinement-like phase, while in the PM phase it decays exponentially at large distances, which is a confinement-like phase. Note that in both phases, the gauge-invariant single-particle correlator always decay exponentially. We also present contour plot of entanglement entropy measured at the central site in Fig.~\ref{fig:matz2}(b), which is defined as $S_E = - \Tr\rho_{L/2} \log \rho_{L/2}$, with $\rho_{L/2} = \Tr_{i=1,2,\cdots,L/2} \dyad{\Psi_{\gs}}$. It is found that the system is more entangled in the PM phase than the AFM phase. The phase transition boundary is also evident from this plot.

\section{\label{sec:discussion-outlook}Summary and outlook}
In this work, we propose to quantum simulate LGT with gauge fixing. In particular, we firstly explain the concept of gauge fixing on the Hamiltonian level. The proper gauge fixing procedure requires the matrix elements of the Hamiltonian of a LGT before and after gauge fixing are identical, as well as the matrix elements of the gauge invariant observable. Thus by quantum simulation LGTs with fixed gauge, one can acquire the full information about the unfixed original models. Usually, the gauge fixed model is much simpler and easier to implement in experiments. Then we consider the simplest LGT, namely the Ising LGT, and discuss in detail on how to fix the gauge for this model on a ladder geometry. After gauge fixing, it becomes a model describing fermions hopping on a ladder subject to a fluctuating dynamical flux. We then provide a Floquet engineering scheme to simulate the corresponding gauge-fixed Hamiltonian. Lastly, we study this gauge-fixed Hamiltonian in the limiting cases and use DMRG to investigate various gauge-invariant correlators and many-body phase diagram. There are basically two phases identified, one is the antiferromagnetic phase, where the Wilson loop operator quickly converges to a finite value and does not decay at large distance, reminiscent of the deconfinement phase of the unfixed model. The other is the paramagnetic phase where the Wilson loop operator decays exponentially at large distance, and resembles the confinement phase in the unfixed model.

Here we elaborate in more detail on the comparison of the gauge fixing approach proposed in this work and others. As already mentioned in the introduction, the advantages of gauge fixing are obvious. Because it reduces size of the Hilbert space, it naturally simplifies the physical setup for analog quantum simulations in the sense that the physical degrees of freedom involved can be reduced.
Moreover, since the strict local gauge constraint is bypassed, one does not have to come up with some sophisticated mechanisms to implement them, which is usually the most difficult part of other approaches for quantum simulating LGTs~\cite{schweizer2019,yang2020,mil2020}.
There are nonetheless disadvantages for this approach.
For example, since gauge fixing enforces local gauge constraint among different lattice sites, it generally increases degree of non-locality of the system, such as $\sigma^x_{\expval{j,j+1},\alpha}$ living on legs of the ladder, i.e., Eq.~\eqref{sigmaxonleg}, which becomes non-local after gauge fixing. Systematic studies comparing these different approaches are therefore an important future work.

Along the direction of gauge fixing, it is interesting to consider LGT with other more complicated gauge symmetry group like U(1) or SU(2). It is expected that the gauge-fixed Hamiltonian for these models should be much simpler than the original ones for quantum simulations. Note that the specific experimental scheme adopted is not restrict to Floquet engineering. One should design suitable proposals based on the given gauge-fixed Hamiltonian. We leave it for future studies to consider these more sophisticated cases.

\begin{acknowledgements}
J. Wang is supported by the Fundamental Research Funds for the Central Universities. W. Zheng is supported by NSFC (Grants No.~GG2030007011 and No.~GG2030040453) and Innovation Program for Quantum Science and Technology (Grants No.~2021ZD0302004).
We thank the HPC-ITP for the technical support and generous allocation of CPU time.
\end{acknowledgements}

%

\end{document}